\documentclass[aps,nofootinbib,floatfix,preprint,12pt]{revtex4} 
\usepackage{graphicx}
\newcommand{\sfig}[2]{
\includegraphics[width=#2]{#1}
        }
\newcommand{\Sfig}[2]{
    \begin{figure}[thbp]
    \sfig{#1.eps}{.9\columnwidth}
    \caption{\small {#2}}
    \label{fig:#1}
    \end{figure}
}

\newcommand{\rf}[1]{\ref{fig:#1}}

\def\be{\begin{equation}}
\def\ee{\end{equation}}
\def\bea{\begin{eqnarray}}
\def\eea{\end{eqnarray}}

\begin{document}
\title{The Real Problem with MOND\footnote{Honorable Mention, Gravity Research Foundation 2011 Awards}}
\author{Scott Dodelson$^{1,2,3,\dagger}$}        
\affiliation{$^1$Center for Particle Astrophysics, Fermi National Accelerator Laboratory, Batavia, IL~~60510}
\affiliation{$^2$Department of Astronomy \& Astrophysics, The University of Chicago, Chicago, IL~~60637}
\affiliation{$^3$Kavli Institute for Cosmological Physics, Chicago, IL~~60637}
\date{\today}          
\begin{abstract}
Gravitational potentials in the cosmos are deeper than expected from observed visible objects, a phenomenon usually attributed
to dark matter, presumably in the form of a new fundamental particle. Until such a particle is observed, the jury remains out on dark matter, and modified gravity models must be considered. The class of models reducing to MOdified Newtonian Dynamics (MOND) in the weak field limit does an excellent job fitting the rotation curves of galaxies, predicting the relation between baryonic mass and velocity in gas-dominated galaxies, and explaining the properties of the local group. Several of the initial challenges facing MOND have been overcome, while others remain. Here I point out the most severe challenge facing MOND.
\end{abstract}
\maketitle

\newcommand\aap{Astron. \& Astrophys.}
\newcommand\apjl{Astrophys. J. Lett.}

\newpage

A wide variety of observations have shown that gravitational potentials in the cosmos are deeper than would be expected if these were produced by the matter radiating light. The classic example is the set of flat rotation curves of galaxies, far from the apparent concentration of mass. On larger scales, galaxy clusters have dynamical masses much larger than those that can be explained by the gas or the individual galaxies. Indeed, both galaxies
and galaxy clusters deflect light as if they contained much more matter than is visible. On the largest scales, the map of the cosmic microwave background (CMB) shows that inhomogeneities in the photon-baryon plasma peaked at about one part in 10,000 when the Universe was 380,000 years old. General relativity makes a firm prediction that the amplitude of these inhomogeneities has grown by a factor of one thousand since then, a prediction that is embarrassingly wrong coming from a species that owes its existence to nonlinear structures. The seemingly unavoidable conclusion is that the early CMB maps do {\it not} trace all the matter -- only the baryonic component -- and, in fact, the hidden matter is much more inhomogeneous than the visible component. 

All of these observations point to the existence of {\it dark matter}, in the form of a new elementary particle never before produced in accelerators. Not only has dark matter never been observed in accelerators, it has also not been seen in direct detection experiments (in which the recoil energy of a nucleus impacted by a dark matter particle is observed) or in indirect detection experiments (in which the debris from dark matter annihilations in space are observed). The only evidence to date for dark matter comes from its gravitational effects. Until that situation changes, modified gravity will remain a viable alternative to dark matter.

One of the first, and still one of the most successful, models that explains the effects usually attributed to dark matter is MOdified Newtonian Dynamics (MOND)~\cite{Milgrom}. The initial version proposed that the gravitational acceleration $a$ caused by a point mass is Newtonian ($MG/r^2$) as long as $a\gg a_0$, since determined~\cite{mcgaugh} to be $1.2\times 10^{-8}$ cm$^2$ sec$^{-1}$. In the regions where $a\ll a_0$, the acceleration is instead the geometric mean of the Newtonian value and $a_0$. This immediately explains flat rotation curves, the Tully-Fisher relation~\cite{mcgaugh}, and many observations of the local Universe~\cite{benoit}. It also re-opens an age-old debate when scientists are confronted with gravitational anomalies: Do the fundamental laws of nature need to be changed or do we assume that the laws are correct but some form of matter is undiscovered. The most dramatic examples of this debate are the discovery of Neptune (dark matter) and the anomalous precession of the perihelion of Mercury (modifed gravity). It is quite likely that the latest incarnation of this debate will be decided in the next decade.

MOND faces numerous challenges, many of which were identified early on; some of which have been resolved or at least ameliorated; and others that have become more serious. It was quickly realized~\cite{1984ApJ...286....7B} that the charge that MOND is not a relativistically covariant theory can be easily accommodated by embedding it in a scalar-tensor model of gravity. Because the Einstein and Jordan metrics are related 
by a conformal factor, scalar-tensor models predict that photons (with invariant distance $ds^2=g_{\mu\nu}dx^\mu dx^\nu=0$ independent of the conformal factor) experience the same deflection as in pure general relativity. This prediction, coupled with the hypothesis of no dark matter, is incompatible with data from gravitational lensing studies, which demonstrate conclusively that light paths are distorted even if they pass far from the visible regions of galaxies. Bekenstein~\cite{Bekenstein:2004ne} proposed breaking the conformal relation of the two metrics with a vector field $A_\mu$ so that
\begin{equation}
g^{\rm Jordan}_{\mu\nu} = e^{-2\phi} g^{\rm Einstein}_{\mu\nu} - 2 A_\mu A_\nu\sinh(2\phi)
\end{equation}
where $8\pi G=1$. The new Tensor-Vector-Scalar model, TeVeS, does indeed produce additional lensing far from the centers of galaxies.

MOND does not explain well the dynamics of galaxy clusters, and that tension was increased by the detailed study~\cite{2006ApJ...648L.109C} of the Bullet Cluster, which seems to require collisionless dark matter. The arguments in defense of MOND/TeVeS -- we should not draw conclusions from a single object, Cold Dark Matter models also have difficulty accounting for this very unusual cluster~\cite{Lee:2010hja}, and the full calculation for an axisymmetric lens has not been obtained so far in TeVeS -- are true but unconvincing, reflecting the reality that clusters remain a challenge for any no-dark matter model.

The growth of structure argument can finally be confronted given a concrete theory, such as TeVeS. Does TeVeS, without dark matter, allow perturbations to grow from a part in ten thousand to greater than unity today? The answer is ``yes,''~\cite{Dodelson:2006zt} and for a reason that is surprisingly compelling. Perturbations to the vector field, introduced by Bekenstein for completely different reasons, drives a difference between the two scalar gravitational potentials, which in turn produces anomalous growth in the overdensities. The power spectrum, which quantifies the clumpiness of structure, is therefore enhanced as depicted in Fig.~\rf{power}. Normalizing at the last scattering redshift of the CMB, the dimensionless quantity, $k^3P(k)/2\pi^2$, which is greater than unity on scales that have gone nonlinear, can indeed exceed one in TeVeS, so there is enough time for nonlinear structures to form in the Universe. The vector field then solves the no-structure problem. This could be a coincidence: perhaps the vector field just happens to play a dual role solving two of the most vexing problems facing a no-dark matter model, or it could be a sign that vector fields are an integral part of the solution.

\Sfig{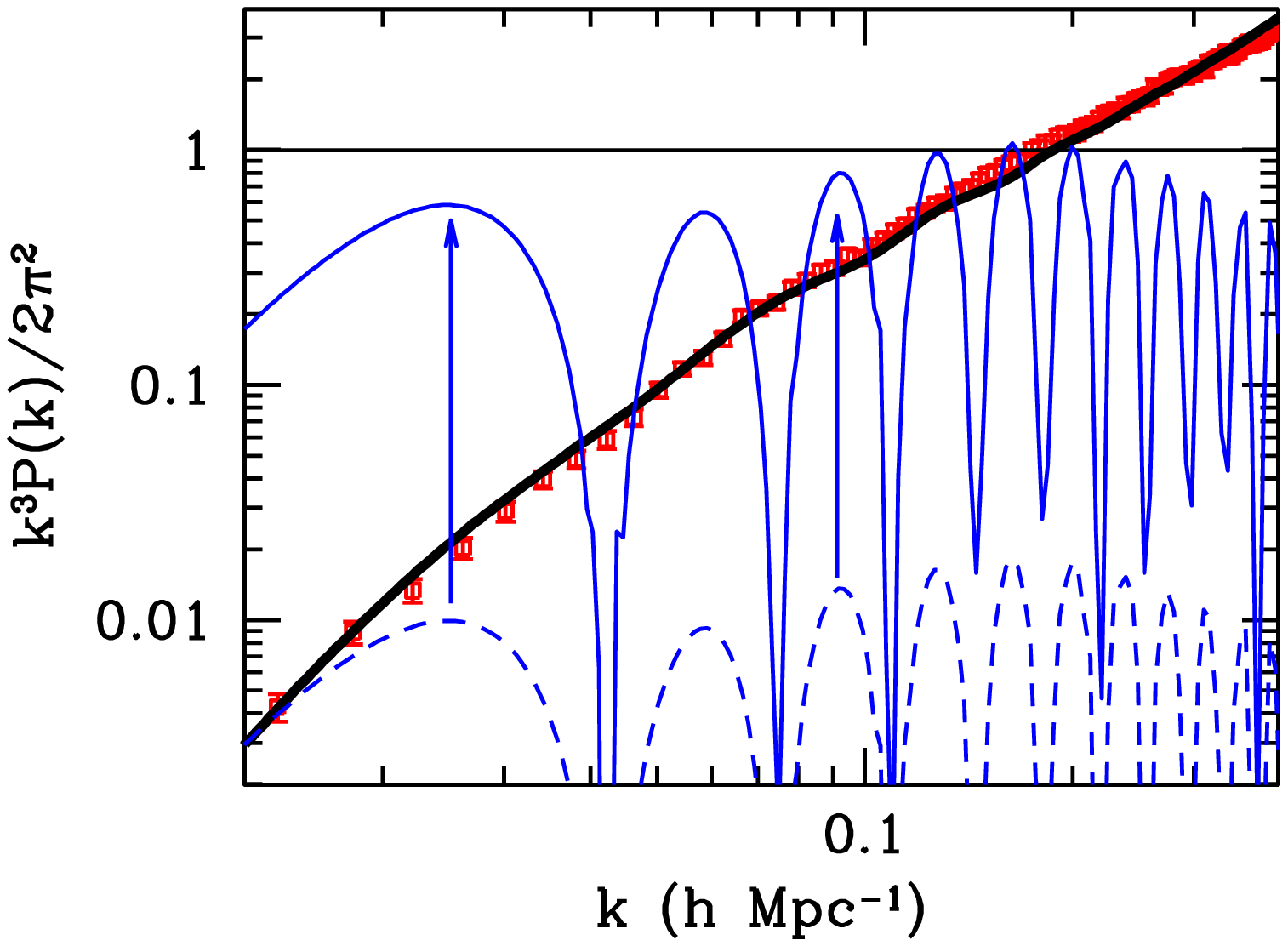}{The power spectrum of matter. Red points with error bars are the data from the Sloan Digital Sky Survey~\cite{Percival:2006gt}; heavy black curve is the $\Lambda$CDM model, which assumes standard general relativity and contains 6 times more dark matter than ordinary baryons. The dashed blue curve is a ``No Dark Matter'' model in which all matter consists of baryons (with density equal to 20\% of the critical density), and the baryons and a cosmological constant combine to form a flat Universe with the critical density. This model predicts that inhomogenities on all scales are less than unity (horizontal black line), so the Universe never went nonlinear, and no structure could have formed. TeVeS (solid blue curve) solves the no structure problem by modifying gravity to enhance the perturbations (amplitude enhancement shown by arrows). While the amplitude can now exceed unity, the spectrum has pronounced Baryon Acoustic Oscillations, in violent disagreement with the data.}

The importance of the vector field is the second coincidence/miracle associated with TeVeS. The first concerns the numerical value of the fundamental acceleration parameter, $a_0$. It is approximately equal to $H_0$, the current value of the Hubble expansion parameter. The striking aspect of this relates to the recent discovery of the acceleration of the Universe. Models introduced to explain the acceleration either introduce a new substance, dark energy, or modify general relativity. In either case, the fundamental action requires a new dimensionful parameter, and this parameter is inevitably of order $H_0$. This too could be a coincidence: perhaps these two completely different sets of observations -- those typically associated with dark matter and those associated with dark energy -- by chance require modifications to standard physics at the same scale, or this could be pointing to a new fundamental mass scale in physics.

Anisotropies in the CMB also provide mixed results for MOND/TeVeS. Early indications of a low second peak were consistent with general arguments about MOND, and indeed the current value of the ratio of the heights of the first and second acoustic peaks is consistent with a no-dark matter model. The problem is that the third peak should be very small in a baryon-dominated model that lacks the extra gravitational forcing usually supplied by dark matter. The observed third peak is therefore very difficult to accommodate in the framework of MOND. No recent fits have been carried out varying all relevant parameters, but there was tension even back in 2003~\cite{McGaugh:2003qw} before the higher peaks had been measured accurately. Escaping these constraints presumably requires a combination of primary and secondary anisotropies conspiring to fit the data on small scales.

The biggest challenge facing MOND today is the shape of the matter power spectrum.
The shape depicted in Fig.~\rf{power} is related to the acoustic oscillations observed in the CMB. If the Universe is dominated by dark matter, these matter oscillations, dubbed Baryon Acoustic Oscillations (BAO), are highly suppressed as the baryons fall into the potential wells created by dark matter, leaving only percent level traces of the primordial oscillations. In a no-dark matter model, on the other hand, the oscillations should be just as apparent in matter as they are in the radiation. Indeed, Fig.~\rf{power} illustrates that -- even if a generalization such as TeVeS fixes the amplitude problem -- the shape of the predicted spectrum is in violent disagreement with the observed shape.

\bibliography{mond}
\end{document}